\documentclass[%
reprint,
superscriptaddress,
amsmath,amssymb,
aps,
prr,
]{revtex4-2}
\usepackage[english]{babel}
\usepackage[table]{xcolor}
\definecolor{lightgray}{gray}{0.9}
\usepackage{epsfig}
\usepackage{graphicx}
\usepackage[utf8]{inputenc}
\usepackage[T1]{fontenc}
\usepackage{scalefnt}
\usepackage{multirow}
\usepackage{float}
\usepackage{appendix}

\begin{document}
	
	% \preprint{APS/123-QED}
	\preprint{APS/123-QED}
	\title{YNiSn$_2$: A candidate Dirac semimetal}% Force line breaks with \\
	
\author{G. S. Freitas}
\email{gsfreitas@lanl.gov}
\affiliation{Instituto de F\'isica Gleb Wataghin, UNICAMP, 13083-859, Campinas, SP, Brazil.}
\affiliation{Los Alamos National Laboratory, Los Alamos, New Mexico 87545, USA.}
\author{C. Kaufmann Ribeiro}
\affiliation{Los Alamos National Laboratory, Los Alamos, New Mexico 87545, USA.}
\author{F. B. Carneiro}
\affiliation{Los Alamos National Laboratory, Los Alamos, New Mexico 87545, USA.}
\affiliation{Brazilian Center for Research in Physics, Rio de Janeiro-RJ, Brazil.}

\author{H. Pizzi}
\affiliation{Instituto de F\'isica Gleb Wataghin, UNICAMP, 13083-859, Campinas, SP, Brazil.}
\author{R. C. Santos}
\affiliation{Instituto de F\'isica Gleb Wataghin, UNICAMP, 13083-859, Campinas, SP, Brazil.}
\author{D. L. Passos}
\affiliation{Instituto de F\'isica Gleb Wataghin, UNICAMP, 13083-859, Campinas, SP, Brazil.}
\author{M. E. M. Ignacio}
\affiliation{Instituto de F\'isica Gleb Wataghin, UNICAMP, 13083-859, Campinas, SP, Brazil.}
\author{K. R. Pakuszewski}
\affiliation{Instituto de F\'isica Gleb Wataghin, UNICAMP, 13083-859, Campinas, SP, Brazil.}
\affiliation{Univ. Montreal, Dept. Phys., Montreal, PQ H2V 0B3, Canada.}
\author{A. P. Machado}
\affiliation{Instituto de F\'isica Gleb Wataghin, UNICAMP, 13083-859, Campinas, SP, Brazil.}
\author{F. S. Oliveira}
\affiliation{Instituto de F\'isica Gleb Wataghin, UNICAMP, 13083-859, Campinas, SP, Brazil.}
\author{M. M. Piva}
\affiliation{Max Planck Institute for Chemical Physics of Solids, N{\"o}thnitzer Str. 40, Dresden, Germany}
\author{E. M. Bittar}
\affiliation{Brazilian Center for Research in Physics, Rio de Janeiro-RJ, Brazil.}
\author{C. Adriano}
\affiliation{Instituto de F\'isica Gleb Wataghin, UNICAMP, 13083-859, Campinas, SP, Brazil.}
\affiliation{Univ. Montreal, Dept. Phys., Montreal, PQ H2V 0B3, Canada.}
\author{Y. Kopelevich}
\affiliation{Instituto de F\'isica Gleb Wataghin, UNICAMP, 13083-859, Campinas, SP, Brazil.}
\author{F. Ronning}
\affiliation{Los Alamos National Laboratory, Los Alamos, New Mexico 87545, USA.}
\author{J. D. Thompson}
\affiliation{Los Alamos National Laboratory, Los Alamos, New Mexico 87545, USA.}
\author{S. M. Thomas}
\affiliation{Los Alamos National Laboratory, Los Alamos, New Mexico 87545, USA.}
\author{P. F. S. Rosa}
\affiliation{Los Alamos National Laboratory, Los Alamos, New Mexico 87545, USA.}
\author{P. G. Pagliuso}
\email{pagliuso@unicamp.br}
\affiliation{Instituto de F\'isica Gleb Wataghin, UNICAMP, 13083-859, Campinas, SP, Brazil.}
\affiliation{Los Alamos National Laboratory, Los Alamos, New Mexico 87545, USA.}

	\date{\today}
	
	\begin{abstract}
    
 We report the synthesis and physical properties of the new compound YNiSn$2$, which crystallizes in the orthorhombic \textit{Cmcm} structure. The material exhibits semimetallic behavior and develops a giant positive magnetoresistance approaching 1200\% at $B = 16$ T. Pronounced de Haas--van Alphen and Shubnikov--de Haas oscillations reveal a dominant quasi-two-dimensional Fermi surface with an exceptionally small cyclotron effective mass of $m^{*} = 0.08~m{0}$, indicating light carriers and a tiny Fermi surface pocket. The strong anisotropy revealed by Shubnikov–de Haas quantum oscillation measurements highlights the low-dimensional electronic character of YNiSn$_2$, positioning it as a promising Dirac semimetal candidate.

	\end{abstract}

	\maketitle
	
	\section{\label{sec:intro}Introduction}

    Nontrivial topology in condensed matter physics has attracted significant interest over the past decades, driven by the discovery of novel quantum phases such as two- and three-dimensional topological insulators (TIs), excitonic insulators (EIs), as well as Dirac and Weyl semimetals (DSMs and WSMs) \cite{bernevig_quantum_2006,konig_quantum_2007,culcer_transport_2020,hasan_colloquium_2010,yan_topological_2017,zhang_towards_2018,armitage_weyl_2018,jerome_excitonic_1967,kohn_excitonic_1967,blatt_bose-einstein_1962,kotov_electron-electron_2012,kunes_excitonic_2015,jia_evidence_2022}. These materials host low-energy quasiparticles that mimic relativistic Dirac or Weyl fermions, arising from symmetry-protected band crossings near the Fermi level. While Weyl semimetals are inherently topological, the topological character of Dirac semimetals can depend sensitively on symmetry constraints and band inversion details. DSMs are characterized by three-dimensional linear band crossings near the Fermi level, where conduction and valence bands intersect at discrete Dirac nodes \cite{yan_topological_2017,armitage_weyl_2018}. This linear dispersion leads to high carrier mobility, large and often non-saturating magnetoresistance, and unusual magnetotransport phenomena, including signatures associated with the chiral anomaly.

    The search for materials hosting unconventional semimetallic behavior remains central to understanding emergent electronic phases in topological and correlated systems. Here, we investigate the physical properties of YNiSn$_2$ single crystals, crystallizing in the orthorhombic \textit{Cmcm} (No. 63) structure. While displaying semimetallic behavior at low fields, YNiSn$_2$ develops a giant positive magnetoresistance approaching 1200\% at $B = 16$ T, accompanied by a sublinear field dependence.

    Furthermore, pronounced quantum oscillations uncover a dominant quasi-two-dimensional Fermi surface in YNiSn$_2$ at low temperatures. Well-resolved de Haas–van Alphen (dHvA) oscillations observed for magnetic field applied along the $b$ axis yield an exceptionally small cyclotron effective mass of $m^{*} = 0.08~ m_{0}$, indicative of very light carriers and a remarkably small Fermi surface cross section. The corresponding oscillation frequency implies a tiny pocket at the Fermi level, consistent with a nodal-like electronic structure and pronounced anisotropy of the underlying bands. Complementary Shubnikov–de Haas (SdH) analysis independently supports this quasi-two-dimensional scenario, further evidencing that a 2D-like Fermi surface governs the charge transport at low temperatures. The consistency between thermodynamic (dHvA) and transport (SdH) probes reinforces the robustness of this picture.

    Electronic systems with quasi-two-dimensional Fermi surfaces are of particular interest because reduced dimensionality enhances anisotropic transport and amplifies correlation effects \cite{armitage_weyl_2018,changNodalline2025,wangInterplay2017}. In such systems, restricted phase space and weakened screening can promote interaction-driven instabilities, while Landau quantization under magnetic field becomes more pronounced due to the confined electronic dispersion. Quasi-2D electronic structures often arise in proximity to symmetry-protected band crossings, including Dirac and nodal-line semimetals, where small Fermi surface pockets may emerge from linearly dispersing bands near the Fermi level \cite{changNodalline2025}. These features render quasi-2D semimetals particularly susceptible to magnetic-field-driven electronic reconstruction, making them compelling platforms for realizing Dirac-like quasiparticles. Within this context, the pronounced quasi-two-dimensional electronic structure of YNiSn$_2$ positions it as a promising candidate to explore low-dimensional Dirac semimetal physics and interaction-enhanced instabilities.

	\section{\label{sec:experiment}Experiment}
	
	Single crystalline samples of YNiSn$_{2}$ were grown by the Sn-flux method. High-purity elements of Y, Ni, and Sn, with the stoichiometry 1:1:20, were placed inside an alumina crucible and sealed in an evacuated quartz tube. The sample was heated in a furnace at a rate of 100 $^{\circ}$C/h up to 1100 $^{\circ}$C and kept at this temperature for 12 h. In sequence, the tube was slowly cooled at 2 $^{\circ}$C/h to  900 $^{\circ}$C, followed by a faster cooldown to 550 $^{\circ}$C. Then, the batches were removed from the furnace and placed into a centrifuge to separate the crystals from the flux. This growth method yielded nice platelet-like single crystals with typical 0.5 mm $\times$ 0.5 mm $\times$ 0.1 mm dimensions. Moreover, the synthesized phase stoichiometry for our YNiSn$_{2}$ single crystals was determined to be very close to the 1-1-2 (within 3\%) by elemental analysis using energy dispersive x-ray spectroscopy (EDS). 
	
	Magnetization measurements were performed in a Quantum Design magnetic properties measurement system (MPMS3) equipped with a 7 T magnet. Specific heat measurements were done in a small-mass calorimeter system that employs a quasi-adiabatic thermal relaxation technique. In-plane electrical resistivity was obtained in a commercial AC-frequency bridge using the standard four-probe technique and with the current applied in the crystal plates. These measurements were performed in a commercial physical property measurement system (PPMS) cryostat of Quantum Design with a 16 T magnet. Powder and single-crystal x-ray diffraction were performed using a Panalytical Empyrean powder diffractometer with Cu-K$\alpha$ radiation and a Bruker D8 Venture single-crystal diffractometer equipped with Mo-K$\alpha$ = 0.71073 \AA\ radiation, respectively. The crystallographic orientation of the single crystals was determined using a real-time Laue X-ray system.

	\section{Results and discussion}

    \subsection{Crystalline structure, zero field properties, and susceptibility}
\begin{figure}[!ht]
		\includegraphics[width=0.9\columnwidth]{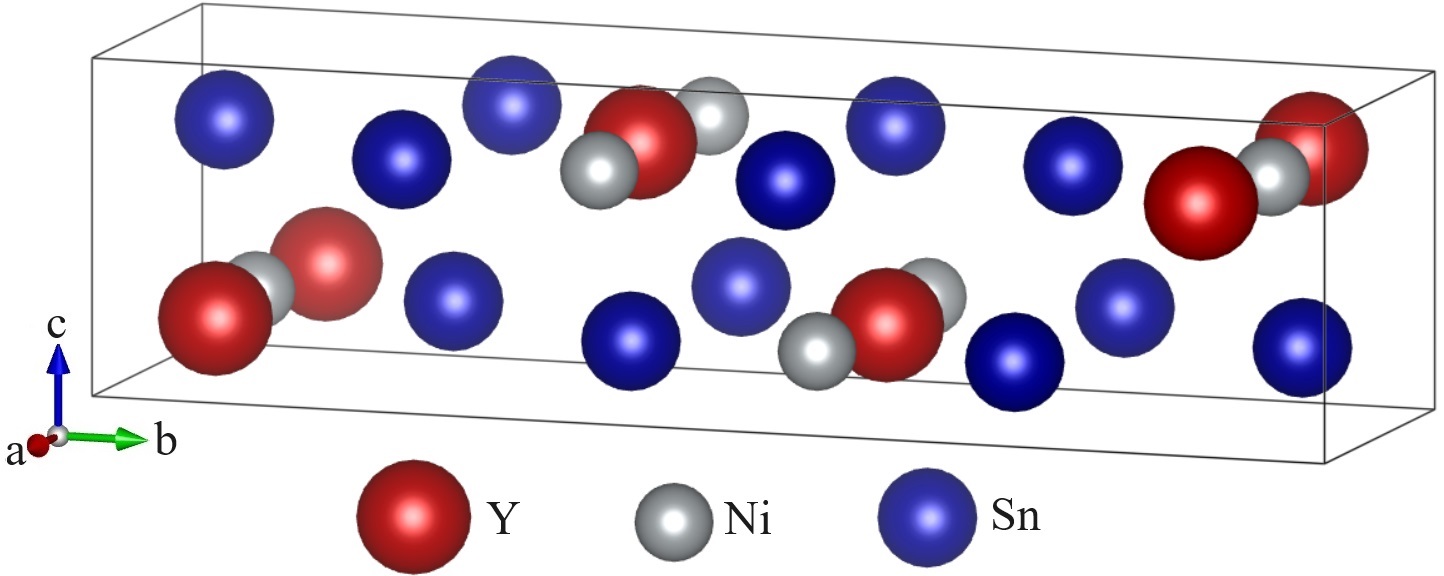}
		\caption{Crystalline structure of YNiSn$_{2}$.}
		\label{estrutura_YNiSn2}
	\end{figure}

    	Room-temperature x-ray powder and single-crystal diffraction measurements confirm that YNiSn$_2$ crystallizes in the orthorhombic \textit{Cmcm} (No. 63) structure with lattice parameters \textit{a} = 4.409 \AA, \textit{b} = 16.435 \AA, and \textit{c} = 4.339 \AA. Figure \ref{estrutura_YNiSn2} displays the obtained crystalline structure. The \textit{CmCm} phase is in contrast with the \textit{Pnma} orthorhombic crystal structure proposed for YNiSn$_{2}$ in the Materials Project Initiative (mp-21981) and previously found for polycrystalline samples of YNiSn$_{2}$\cite{romaka_interaction_2012,sebastian_stannides_2007}. In fact, our results are consistent with the structure previously found for polycrystalline of \textit{R}Ni$_{1-x}$Sn$_{2}$ (\textit{R} = La, Ce, Pr, Nd, Sm) \cite{sebastian_stannides_2007,komarovskaya_crystal_1983,skolozdra_magnetic_1988-1,skolozdra_magnetic_1988} and CeNiSn$_{2}$ \cite{pecharsky_low-temperature_1991}. To assess possible Ni vacancies reported in this compound family, EDS measurements and single-crystal X-ray refinement were performed. Both methods confirm a stoichiometric 1:1:2 composition.

        Interestingly, the centrosymmetric orthorhombic \textit{Cmcm} space group hosts mirror and glide symmetries that can stabilize symmetry-protected band crossings in suitable electronic configurations. Such structural constraints are often conducive to Dirac or nodal-line semimetal behavior \cite{changNodalline2025}, providing a natural framework to discuss such electronic features in YNiSn$_2$.

\begin{figure}[!ht]
		\includegraphics[width=\columnwidth]{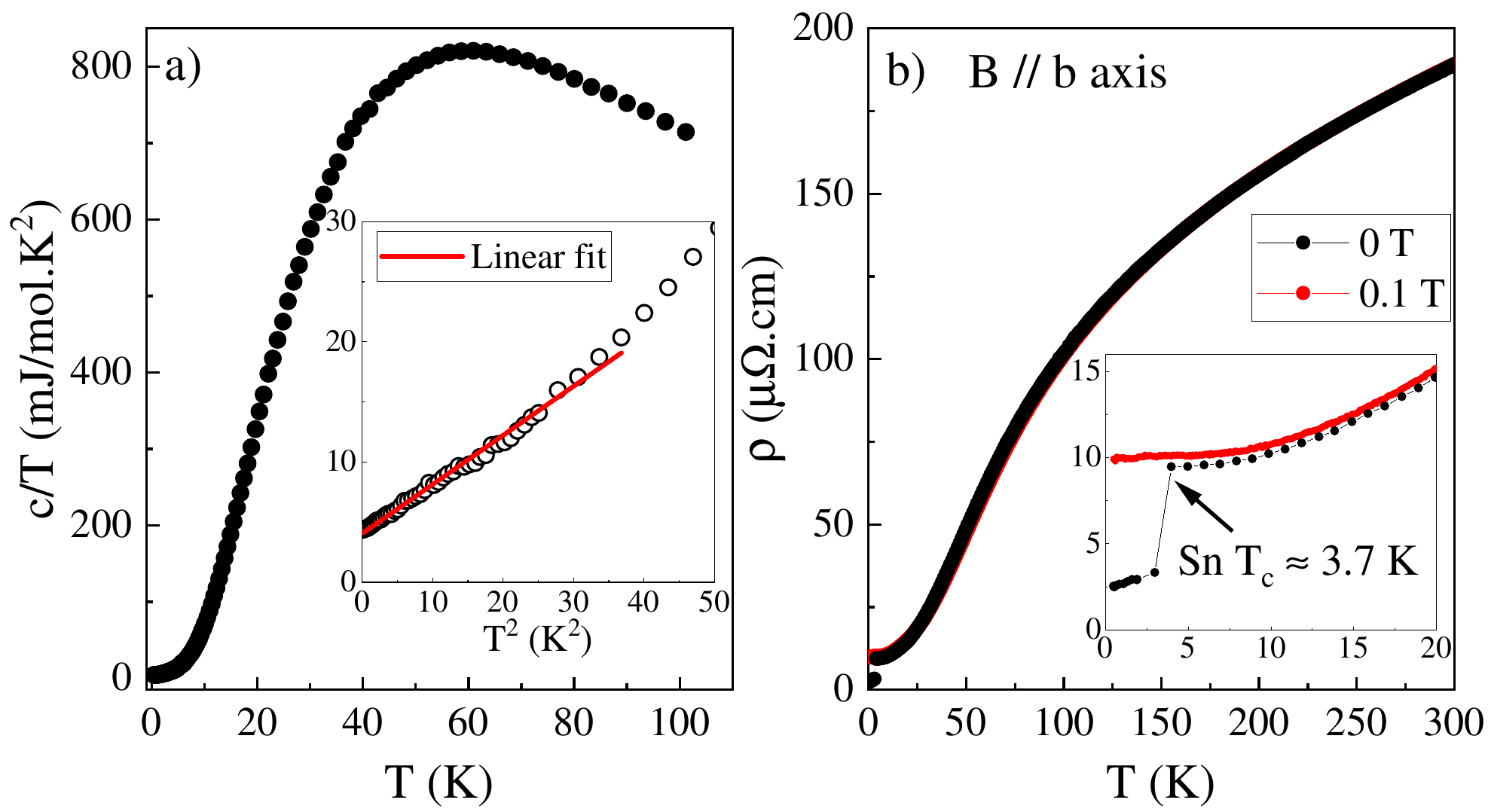}
		\caption{a) Zero-field temperature dependence of the specific heat. The inset shows a linear fit in a $c/T$ vs $T^{2}$ representation.
        b) Electrical resistivity as a function of temperature measured at zero field and under a low magnetic field of 0.1 T applied along the \textit{b} axis. The inset highlights the superconducting transition of residual Sn, which is suppressed under applied field.}
		\label{cp_R_T_YNiSn2}
	\end{figure}

	   Figure \ref{cp_R_T_YNiSn2} presents the zero magnetic field specific heat ($c_{p}$) and electrical resistivity ($\rho$) as a function of temperature for YNiSn$_{2}$. The specific heat data in the temperature range between 0.4 K $<$ T $<$ 6 K was fitted by the expression $c/T = \gamma + \beta T^2$ as seen in the inset of Figure \ref{cp_R_T_YNiSn2}a). The fitting parameters obtained from these data are $\gamma$ = 4.0(5) mJ/mol-K$^2$ and $\beta$ = 0.41(2) mJ/mol-K$^4$, indicating a small density of states. 
	
	Assuming a free-electron gas model for YNiSn$_2$, the Sommerfeld coefficient is given by $\gamma =(2/3)\pi^2 k_B^2 \eta(E_F)$ \cite{kittel_introduction_2005}, where $k_B$ is the Boltzmann constant and $\eta(E_F)$ is the density of states at the Fermi level. Using the obtained $\gamma$, we estimate $\eta(E_F)=0.83(4)$ states/eV mol-spin in the zero-field, low-temperature limit. The Debye temperature was extracted from $\beta = 12\pi^4 N R/(5\theta_D^3)$ \cite{kittel_introduction_2005}, where $N=4$ is the number of atoms per formula unit and $R$ is the gas constant, yielding $\theta_D \approx 170$ K.

	The inset of Figure \ref{cp_R_T_YNiSn2}b) presents a zoomed-in plot of $\rho(T)$ measured under applied magnetic fields of 0 T and 0.1 T  parallel to the \textit{b}-axis, revealing a distinct kink at around 3.7 K. This kink originates from the superconducting transition of residual Sn at 3.72 K \cite{eisenstein_superconducting_1954}.

\begin{figure}[!ht]
		\includegraphics[width=\columnwidth]{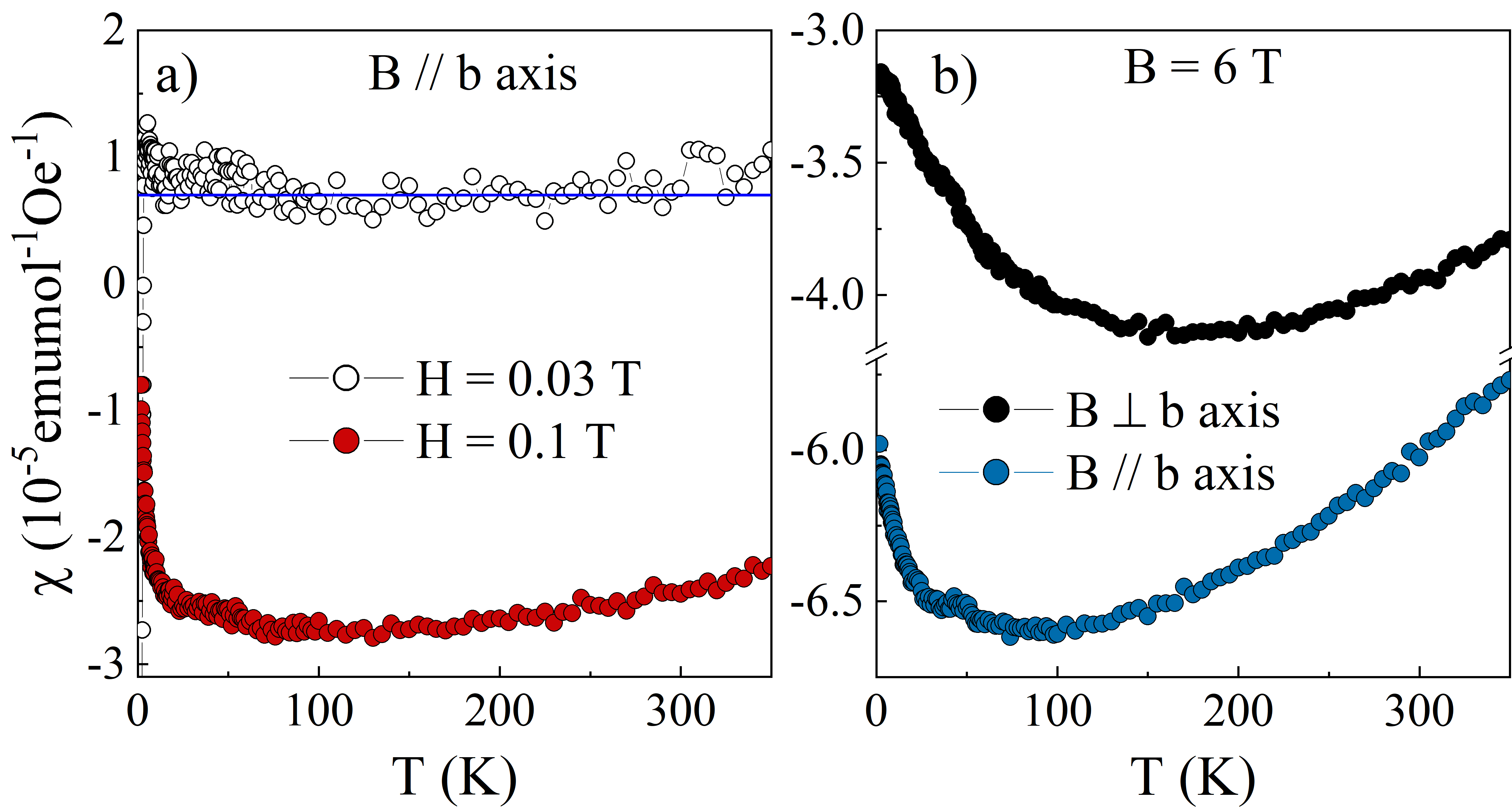}
	\caption{ Temperature-dependent magnetic susceptibility $\chi(T)$ of YNiSn$_2$ single crystals.
    a) $\chi(T)$ measured with magnetic fields of 0.03 T and 0.1 T applied along the \textit{b} axis.
    b) Anisotropic behavior of $\chi(T)$ for a magnetic field of 6 T applied parallel and perpendicular to the \textit{b} axis.}
	\label{chi_YNiSn2}
\end{figure}

The temperature-dependent  magnetic susceptibility $\chi(T)$ of YNiSn$_2$ single crystals is shown in Figure~\ref{chi_YNiSn2}. Overall, the susceptibility exhibits a weak temperature dependence, with the sign and magnitude of $\chi$ depending on the applied magnetic field. Using the density of states $\eta(E_F)=0.83(4)$ states/eV mol-spin extracted from the zero-field specific-heat data in Figure~\ref{cp_R_T_YNiSn2}, the expected electronic spin susceptibility is $\chi_e=2\mu_B^2\eta(E_F)\approx 0.07\times10^{-4}$ emu/mol Oe. This value is in good agreement with the low-field $\chi(T)$ data of YNiSn$_2$ [Figure~\ref{chi_YNiSn2}(a)], indicating that the low-field response is well described by a Pauli-like paramagnetic contribution.

At higher magnetic fields, however, $\chi(T)$ becomes predominantly diamagnetic and shows only weak magnetic anisotropy, as shown in Figure~\ref{chi_YNiSn2}(b). The high-field response is characterized by a broad minimum as a function of temperature, reminiscent of high-mobility semimetals such as NbP and TaP, where similar minima in $\chi(T)$ have been associated with orbital responses from semimetallic bands \cite{leahy_Nonsaturating_2018}. This comparison suggests that the high-field susceptibility of YNiSn$_2$ is dominated by an orbital semimetallic contribution.

\subsection{Magnetization and dHvA analysis}

\begin{figure*}[t]
	\begin{center}
		\includegraphics[width=0.98\textwidth]{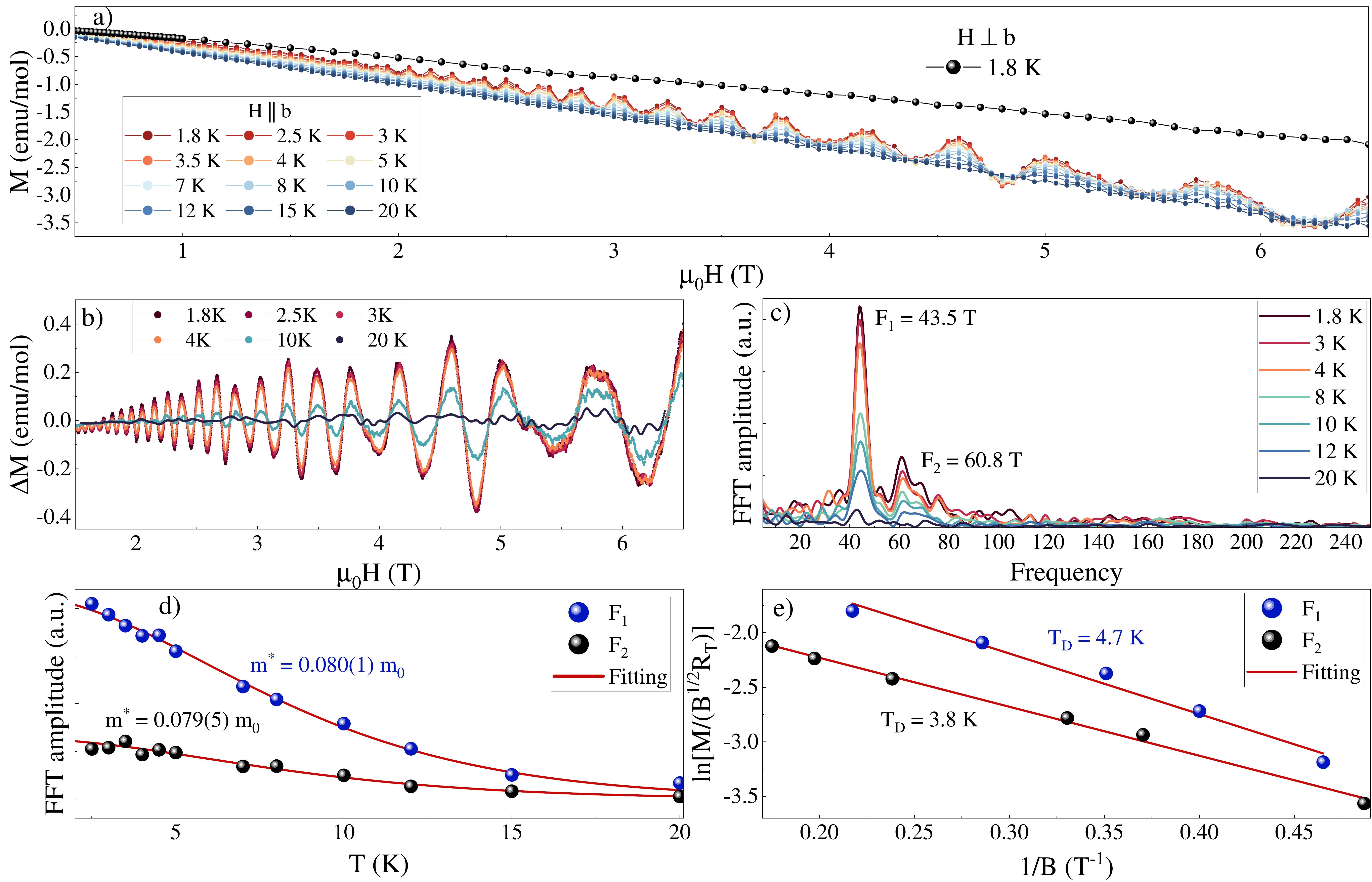}
	\end{center}
	\caption{a) Magnetization $M(B)$ of YNiSn$_2$ for magnetic field applied parallel and perpendicular to the plane, measured up to $B = 6.5$ T at temperatures between 2 and 20 K. b) Oscillatory component $\Delta M$ obtained after subtraction of a linear background. c) Fast Fourier transform (FFT) spectra of $\Delta M$ for temperatures between 2 and 20 K. d,e) Temperature dependence of the oscillation amplitude used to determine the cyclotron effective mass $m^{*}$d) and the Dingle temperature $T_D$ e). } 
	\label{MxH_YNiSn2}
\end{figure*}

Figure \ref{MxH_YNiSn2}a) shows the magnetization versus the magnetic field applied parallel/perpendicular to the \textit{b} axis up to $B$ = 6.5 T for several temperatures ranging from 1.8 to 20 K. The magnetization data exhibit pronounced de Haas–van Alphen (dHvA) oscillations, which are enhanced after subtraction of the high-temperature diamagnetic background [Figure~\ref{MxH_YNiSn2}b)]. The absence of oscillations for magnetic field applied within the \textit{ac} plane supports the presence of a strongly anisotropic Fermi surface. Figure \ref{MxH_YNiSn2}c) presents the fast-Fourier transform (FFT) spectrum in which two very close distinguishable frequencies at $F_1$ = 43.5 T and $F_2$ = 60.8 T were observed. It is important to point out that although several oscillation frequencies for different orientations can be found in the pure Sn, its frequency ranges from 170 T to 10$^5$ T, much larger than the two frequencies observed for YNiSn$_2$ \cite{roger_fermi_1976,finkelstein_fermi_1974}.

Based on the Onsager relation, $F$ = ($\Phi_0$/2$\pi^2$)$A_F$, where $A_F$ is the cross-section area of the Fermi surface normal to the applied magnetic field, we extracted an estimated value of 4.1 X 10$^{-3}$ \AA$^{-2}$ for frequency 43.5 T. This cross-sectional area $A_F$ corresponds to approximately 1\% of the basal-plane area of the Brillouin zone ($\approx 0.44$ \AA$^{-2}$), indicating that the dHvA oscillations originate from a very small Fermi surface pocket. Assuming a circular cross section, the corresponding Fermi wave vector is estimated as $k_F = \sqrt{A_F/\pi} = 0.034$ \AA$^{-1}$.

The nature of the carriers participating in quantum oscillations can be revealed from further quantitative analyses of the dHvA oscillations. If the higher harmonic frequency is not significant, the oscillatory magnetization amplitude ($\Delta M$) can be described using the Lifshitz-Kosevich (LK) formula: \cite{sankar_crystal_2018,shoenberg_magnetic_1984,hu_nearly_2017}
\begin{equation}
	\Delta M \propto B^{1/2}R_TR_D\cos[2\pi(\frac{F}{B}+\gamma-\delta)]  \label{1}
\end{equation}
where 
\begin{equation}
	R_T=\alpha T \mu/[B \sinh(\alpha  T \mu/B)],  \label{2.1}
\end{equation}
\begin{equation}
	R_D=exp(-\alpha T_D \mu/B) \label{3.1}
\end{equation}
wherein $\mu$ = $m^{*}/m_{0}$, with $m_{0}$ is the free electron mass, and $\alpha$ = 2$\pi k_Bm_{0}$/$\hbar e = 14.69$ $T/K$ is a constant.

From fits of the temperature dependence of the FFT amplitude to the thermal damping factor $R_T$ of the Lifshitz–Kosevich formalism [Figure~\ref{MxH_YNiSn2}d)], we extract a cyclotron effective mass of $m^{*} \approx 0.08,m_{0}$ for both $F_1$ and $F_2$. The remarkably small effective mass is consistent with a Fermi surface derived from linearly dispersing bands, as commonly observed in Dirac and Weyl semimetals \cite{armitage_weyl_2018,hu_nearly_2017}. The resulting Fermi velocity, $v_{F} = \hbar k_{F}/m^{*} \approx 6.9 \times 10^{5}$ m/s, lies within the range typically observed in Dirac semimetals, such as ZrSiS \cite{singha_large_2017}, TaAs \cite{sankar_crystal_2018}, BaMnBi$_2$ \cite{li_electron-hole_2016}, TlBiSSe \cite{le_mardele_evidence_2023}, among others. Thereby reinforcing the Dirac-like nature of the low-energy electronic excitations in YNiSn$_2$.

Furthermore, the field dependence of the oscillation amplitude, normalized by the thermal damping factor $R_T$ and fitted to the Dingle factor $R_D$ [Figure~\ref{MxH_YNiSn2}e)], yields Dingle temperatures of $T_D = 4.7$ K and $T_D = 3.8$ K for $F_1$ and $F_2$, respectively. Taking an average value of $T_D \approx 4$ K, the corresponding quantum scattering time, $\tau = \hbar/(2\pi k_B T_D)$, is estimated to be $\sim 3.6 \times 10^{-13}$ s.

The relatively small Dingle temperature and long scattering time reflect low disorder and high crystalline quality, enabling coherent cyclotron motion and the clear resolution of quantum oscillations. Such a long quasiparticle lifetime, together with the ultralight effective mass and small Fermi surface pocket, reinforces the picture of high-mobility, Dirac-like carriers governing the low-energy electronic structure of YNiSn$_2$.

\subsection{Electric resistance temperature dependence }

\begin{figure}[!ht]
	\includegraphics[width=0.98\columnwidth]{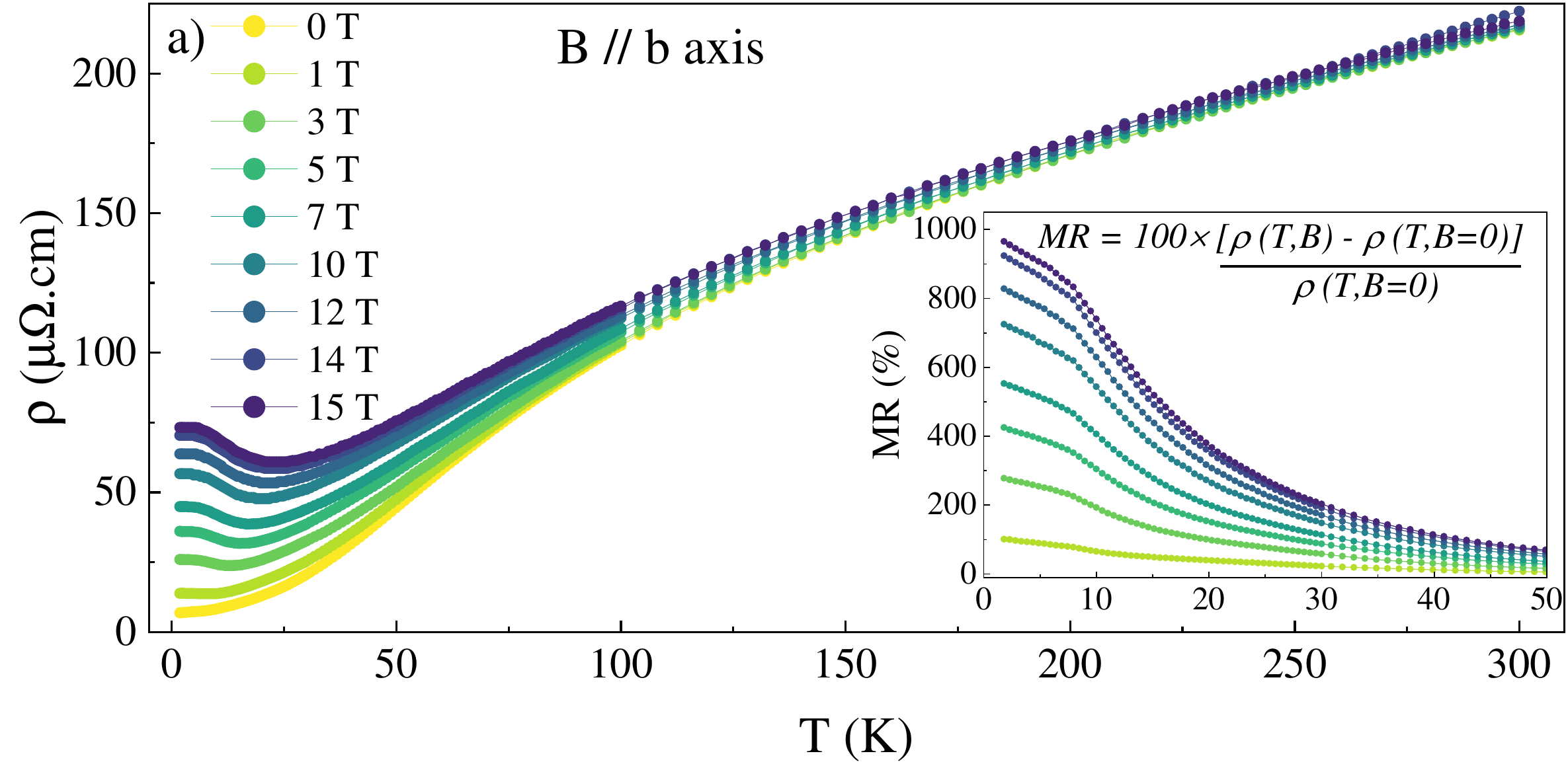}
	\includegraphics[width=0.98\columnwidth]{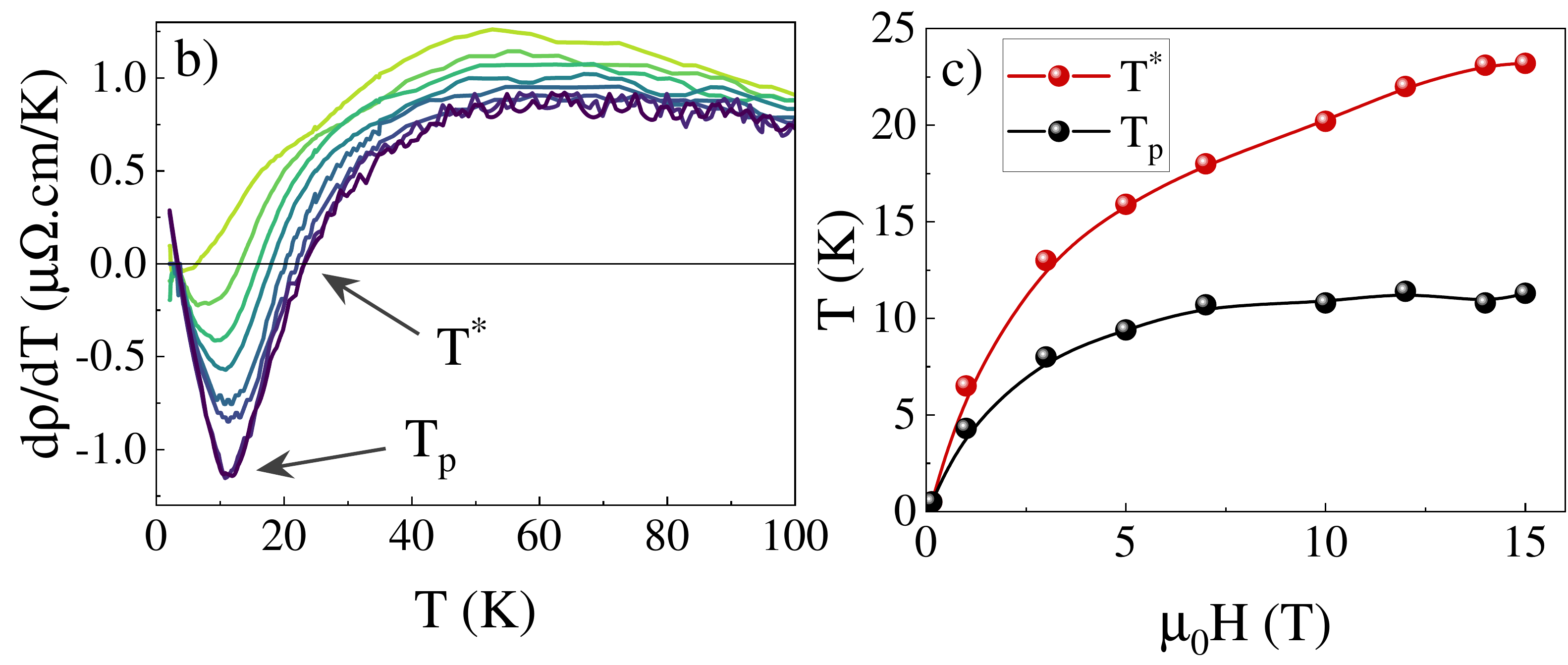}
	\caption{a) In-plane magnetic transport data $\rho(T, B)$ for YNiSn$_2$ with a magnetic field applied perpendicular to the plane. In the inset in a), we present the magnetoresistance (MR) as a function of the temperature. b) The derivative of $\rho(T, B)$ as a function of the temperature, wherein we present the discussed in the text $T^{*}$ and $T_{p}$. c) The behavior to $T^{*}$ and $T_{p}$ as a function of the applied magnetic field. }
	\label{RxT_H_YNiSn2}
\end{figure}

Figure~\ref{RxT_H_YNiSn2}a) displays the in-plane resistivity $\rho(T,B)$ of YNiSn$_2$ for magnetic field applied perpendicular to the \textit{ac} plane. The resistivity exhibits a pronounced field dependence, reaching a positive magnetoresistance of approximately 1000\% at $T = 2$ K and $B = 15$ T [inset of Figure~\ref{RxT_H_YNiSn2}a)]. 

Under applied magnetic field, $\rho(T)$ initially decreases upon cooling, consistent with metallic behavior, before developing a low-temperature minimum followed by an upturn. This behavior suggests that YNiSn$_2$ enters a high-mobility semimetallic regime characterized by $\omega_c \tau = \mu B \gtrsim 1$, where $\omega_c$ is the cyclotron frequency, $\tau$ is the scattering time, and $\mu$ is the carrier mobility \cite{pippard_magnetoresistance_1989}. In this regime, the magnetic field strongly bends carrier trajectories between scattering events, leading to an enhanced magnetoresistance response. This interpretation is also consistent with the high-field diamagnetic susceptibility discussed above, which points to a sizable orbital semimetallic contribution.

Similar field-induced resistivity upturns and large positive magnetoresistance are widely observed in high-mobility semimetals, including WTe$_2$ \cite{ali_Large_2014}, LaSb/LaBi \cite{tafti_resistivity_2016,sun_large_2016}, Cd$_3$As$_2$ \cite{liang_Ultrahigh_2015}, ZrSiS \cite{singha_large_2017}, TaAs \cite{sankar_crystal_2018}, and NbAs$_2$-related dipnictides and monopnictides \cite{peramaiyan_Anisotropic_2018,leahy_Nonsaturating_2018}.

Figure~\ref{RxT_H_YNiSn2}b) presents $d\rho/dT$ derived from the resistivity data in Figure~\ref{RxT_H_YNiSn2}a). The derivative reveals two distinct temperature scales: $T^{*}$, signaling the onset of the crossover into the $w_c*\tau$ $\geq$ 1 regime, and $T_{p}$, corresponding to the inflection point below which a low-temperature resistivity plateau emerges. Both temperature scales increase monotonically with magnetic field [Figure~\ref{RxT_H_YNiSn2}c)]. The temperature of the resistivity minimum, $T^{*}$, increases approximately as $B^{1/2}$. This behavior is consistent with a field-driven crossover into a high-mobility magnetotransport regime, where $\omega_c\tau=\mu B\gtrsim 1$, and resembles the turn-on temperature scaling reported in large-magnetoresistance semimetals such as WTe$_2$ \cite{wang_origin_2015}.

%--------------------------------------------------------------------------------------------------------------

\subsection{Magnetoresistance and SdH}
\begin{figure*}[!ht]
	\includegraphics[width=\textwidth]{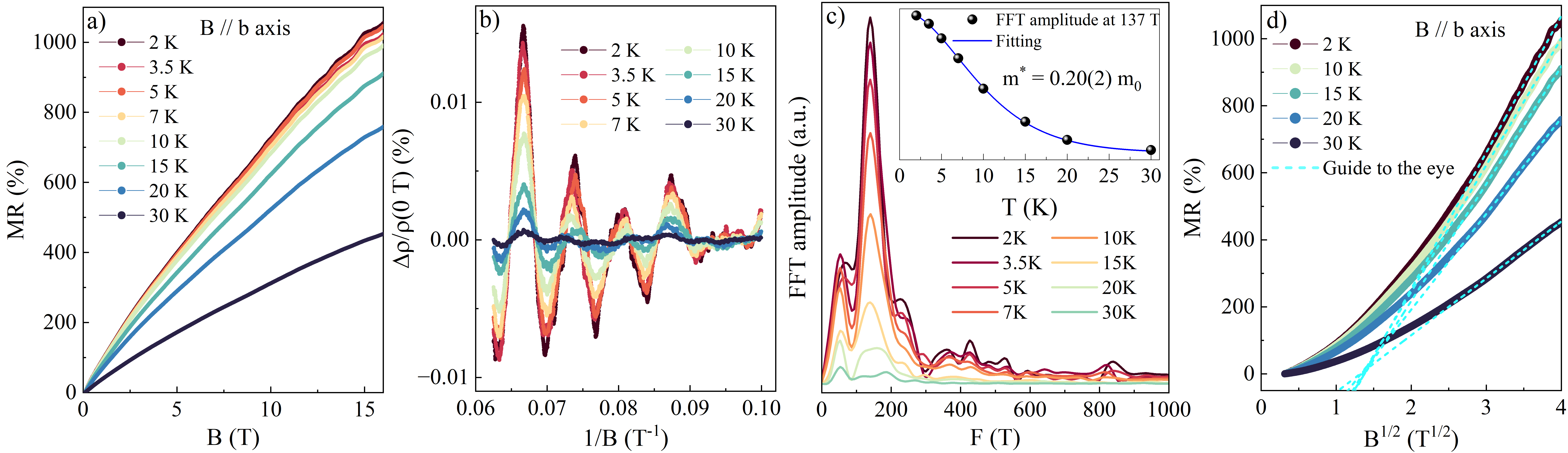}
    \includegraphics[width=\textwidth]{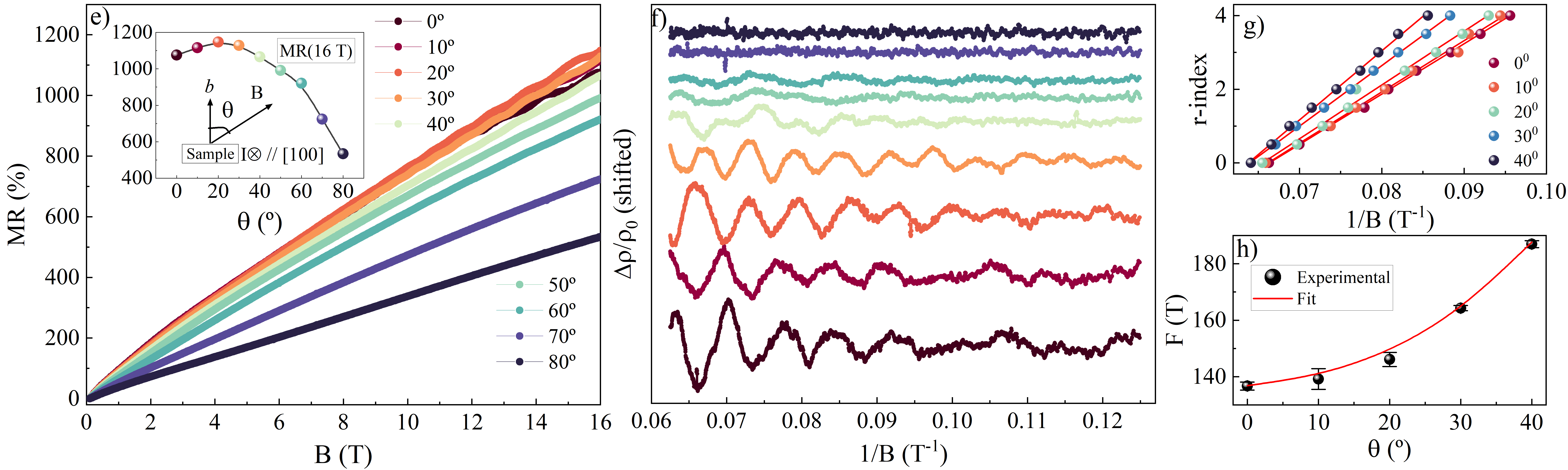}
	\caption{ a) Magnetoresistance as a function of the applied field along the b axis at diferent temperatures. b) Highlight of the SdH oscillations in $\Delta\rho = \rho(H) - y(H)$ (discussed in the text) normalized by $\rho$(0 T) plot as a function of $1/B$ at different temperatures. c) FFT spectrum of the $\Delta\rho/\rho$(0 T) data for several temperatures. d) $MR$ plotted as a function of $B^{1/2}$. Guide lines highlight the crossover between a linear contribution in $B^{1/2}$ and a quadratic ($B$) component. e) $MR$ as a function of the applied magnetic field at different angles with respect to the \textit{b} axis. The inset shows the MR at \(B = 16\) T as a function of the angle. f) $\Delta\rho/\rho$(0 T) as a function of $1/B$ at different temperatures. g) r-index as a function of $1/B$ for different angles of applied field. A linear fit was performed to extract the SdH oscillation frequency. h) Angular dependence of the SdH oscillation frequency with respect to the applied magnetic field direction relative to the \textit{b} axis. The data are fitted with a $F(\theta) \propto 1/\cos\theta$ function.    }
	\label{MRxH_YNiSn2}
\end{figure*}

Figure \ref{MRxH_YNiSn2}a) presents the in-plane magnetoresistance $MR$ = [$\rho(B)$ - $\rho(B = 0)$/$\rho(B = 0)$]$\times$100\%  data for YNiSn$_2$ at different temperatures for the magnetic field applied perpendicular to the \textit{ac} plane. These data revealed a positive magnetoresistance of nearly 1100\% achieved at $T$ = 1.8 K under a magnetic field of $H$ = 16 T. Samples from different batches show a MR ranging from 1000-1400\%. Moreover, Shubnikov-de Haas (SdH) quantum oscillations under high magnetic fields can be observed above 10 T. Although the SdH oscillations exhibit a low signal-to-noise ratio over the accessible field range, a tentative quantum-oscillation analysis was nevertheless performed. In Figure~\ref{MRxH_YNiSn2}b), we show $\Delta\rho = \rho(H) - y(H)$, where $\rho(H)$ is the magnetoresistance and $y(H)$ is a second-order polynomial background fitted between 10 T and 16 T (a polynomial including linear and square-root terms was also tested, yielding similar results).

In the FFT spectrum [Figure~\ref{MRxH_YNiSn2}c)], a prominent peak is observed near 133 T, which is higher than the two frequencies obtained from the dHvA analysis. A weak shoulder appears around 50 T, which would be consistent with the dHvA frequencies. However, given the limited resolution and small oscillation amplitude, we cannot unambiguously confirm the presence of this lower-frequency component. It is important to note that the field windows employed in the SdH and dHvA analyses are different, such that the corresponding FFT spectra are not extracted over the same inverse-field range. Since the accessible frequency content depends on the selected window, this may account for part of the discrepancy between the two measurements. We further note that the 133 T peak is close to three times 43.5 T, raising the possibility that the SdH signal is dominated by a higher harmonic of the fundamental dHvA frequency. High-field measurements with improved signal-to-noise ratio will therefore be essential to disentangle potential harmonic contributions from intrinsic quantum-oscillation frequencies.

In the inset of Figure~\ref{MRxH_YNiSn2}c), we present the thermal damping of the FFT amplitude of the peak at 137~T together with a fit using the LK factor, yielding an effective mass $m^{*}$ = 0.20(2)$m_0$, which is larger than the value obtained from the dHvA analysis. Similar to the frequency discrepancies, the difference in the extracted effective mass may be influenced by the distinct field windows employed in the two analyses. However, it is important to emphasize that SdH measurements are intrinsically more sensitive to scattering effects and therefore probe a transport-weighted effective mass, whereas dHvA measurements reflect the thermodynamic cyclotron mass. The observed difference is thus likely associated with band-dependent scattering.

By fitting MR field dependence with a power law, we found MR $\propto$ $B^{n}$ with $n$ = 0.72(3), close to a square-root dependence. In fact, as presented in Figure \ref{MRxH_YNiSn2}d), a crossover was observed between a low field region displaying a linear dependence and the high field region with a $B^{1/2}$ dependence. Interestingly, a very similar behavior of the MR was theoretically described to appear in certain quasi-2D layered metals \cite{grigoriev_longitudinal_2013}. This could occur when the separation between Landau levels significantly exceeds the width of these levels and the interlayer interactions within the material. The emergence of the square-root dependence of MR is attributed to short-range (point-like) interactions with impurities, leading to changes in the effective scattering rate of electrons when subjected to a magnetic field. This effect is expected in materials where the quantized motion of electrons and their interactions with lattice imperfections are significant \cite{grigoriev_longitudinal_2013, sinchenko_linear_2017}. To the best of our knowledge, no 3D material has been found that presents such MR crossover behavior. Curiously, a very similar theoretical study was performed in disordered graphene, wherein it predicted a square-root MR at high fields \cite{alekseev_strong_2013}. This was observed experimentally in a graphene monolayer \cite{vasileva_magnetoresistance_2019}. For more details, see  references \cite{grigoriev_longitudinal_2013,sinchenko_linear_2017,alekseev_strong_2013,vasileva_magnetoresistance_2019} and references therein.

Figure \ref{MRxH_YNiSn2}e) shows the angular dependence of the MR of the sample. The MR reaches a maximum value of $\approx$ 1200\% at $\theta$ = $ 20^\circ$, where $\theta$ is the angle between the applied magnetic field and the $b$-axis. For $\theta > 20^\circ$, the MR measured at 16~T decreases monotonically, reaching 530\% at $\theta = 80^\circ$. We observe clear Shubnikov--de Haas oscillations from $\theta = 0^\circ$ up to $40^\circ$, which become strongly suppressed for larger angles. 

To extract the oscillation frequency at different angles, we determine the positions of successive peaks and valleys of the oscillatory component and construct a Landau fan diagram by plotting the oscillation index as a function of $1/B$, see Figure \ref{MRxH_YNiSn2}g). Because the available magnetic-field range is not sufficient to reliably assign absolute Landau level indices, we index the oscillations sequentially rather than by absolute Landau level number. The oscillation frequency is then obtained from the slope of a linear fit to the fan diagram.

The oscillation frequency as a function of angle follows the behavior expected for a two-dimensional Fermi surface. According to the Onsager relation, $F = \frac{\hbar}{2\pi e} A_{\mathrm{ext}}$, the frequency is proportional to the extremal cross-sectional area $A_{\mathrm{ext}}$ perpendicular to the applied magnetic field. For a quasi-two-dimensional cylindrical Fermi surface, the projected area scales as $A_{\mathrm{ext}}(\theta) = A_0 / \cos(\theta - \theta_0)$, leading to 
$F(\theta) = \frac{F_0}{\cos(\theta - \theta_0)}$, 
where $F_0 = 138(2)\,\mathrm{T}$ is the frequency for $\mathbf{B} \parallel b$ and $\theta_0 \approx 2^\circ$ accounts for a small sample misalignment.  The excellent agreement with the dependence $1/\cos(\theta)$ indicates that the extremal Fermi surface cross-section in the rotated plane exhibits a two-dimensional character.

\section{Conclusions and Perspectives }

In summary, our study establishes YNiSn$_2$ as a new semimetal with a dominant quasi-two-dimensional electronic structure. Comprehensive characterization of newly synthesized single crystals, including x-ray diffraction, elemental analysis, electrical resistivity, magnetic susceptibility, and specific heat measurements, confirms the high crystalline quality and semimetallic nature of the compound. YNiSn$_2$ exhibits a large, nearly linear magnetoresistance reaching approximately 1200\% at 16 T, consistent with high-mobility carrier dynamics. Well-resolved dHvA oscillations reveal an exceptionally small cyclotron effective mass, $m^{*}=0.08~m_{0}$, together with a small Fermi surface, indicating the presence of light, low-density carriers. In addition, SdH oscillations provide strong evidence for a highly anisotropic, quasi-two-dimensional Fermi surface. Taken together, these results identify YNiSn$_2$ as a promising Dirac semimetal candidate and motivate future spectroscopic and high-field studies to further clarify its electronic structure.

\section{Acknowledgments}

This work was supported by FAPESP (Grants No  2017/10581-1, 2019/26247-9, 2022/09240-3, 2024/00998-6, 2025/03504-7, 2025/19245-0), CNPq (Grants No. 140724/2019-2, 405408/2023-4 and 311783/2021-0), CAPES, and FINEP-Brazil.
F.B.C. Acknowledge the support from CNPq (Proc. 141063/2020-3) and CAPES (Finance Code 001).
We further acknowledge support from the INCT project Advanced Quantum Materials, involving the Brazilian agencies CNPq (Proc. 408766/2024-7), FAPESP (Proc. 2025/27091-3) and CAPES.
The authors would like to acknowledge the LNNano - Brazilian Nanotechnology National Laboratory (LNNano/CNPEM/MCTIC) for providing the equipment and technical support for the experiments involving scanning electron microscopy. 
Work at Los Alamos National Laboratory was performed under the auspices of the U. S. Department of Energy, Office of Science, Basic Energy Sciences, Division of Materials Science and Engineering.

\bibliography{main}

\end{document}